\documentclass[usenatbib,useAMS,usegraphicx]{mn2e}
\usepackage{amsmath}

\def\lesssim{\mathrel{\hbox{\rlap{\hbox{\lower4pt\hbox{$\sim$}}}\hbox{$<$}}}}
\def\gtrsim{\mathrel{\hbox{\rlap{\hbox{\lower4pt\hbox{$\sim$}}}\hbox{$>$}}}}

\title[Searching for the most distant gamma-ray blazars]
{Searching for the most distant blazars with the Fermi Gamma-ray Space Telescope}
\author[Y. Inoue et al.]{Yoshiyuki Inoue,$^1$\thanks{E-mail: yinoue@kusastro.kyoto-u.ac.jp (YI)} Susumu Inoue,$^2$ Masakazu A. R. Kobayashi,$^3$ Tomonori Totani,$^1$
\newauthor
Jun Kataoka$^4$, \& Rie Sato$^5$\\
$^1$Department of Astronomy, Kyoto University, Oiwake-cho, Kitashirakawa, Sakyo-ku, Kyoto 606-8502, Japan\\
$^2$Department of Physics, Kyoto University, Oiwake-cho, Kitashirakawa, Sakyo-ku, Kyoto 606-8502, Japan\\
$^3$Optical and Infrared Astronomy Division, National Astronomical Observatory of Japan, Mitaka, Tokyo 181-8588, JAPAN\\
$^4$Waseda University, 1-104 Totsukamachi, Shinjuku-ku, Tokyo 169-8050, Japan\\
$^5$Department of High Energy Astrophysics, Institute of Space and Astronautical Science (ISAS), Japan Aerospace Exploration Agency (JAXA), \\
3-1-1 Yoshinodai, Sagamihara, 229-8510, Japan}

\begin{document}
\pagerange{\pageref{firstpage}--\pageref{lastpage}} \pubyear{----}

\maketitle

\label{firstpage}
\begin{abstract}
We investigate the prospects for discovering blazars at very high redshifts ($z \gtrsim 3-6$) with the Fermi Gamma-Ray Space Telescope ({\it Fermi}), employing a model for the evolving gamma-ray luminosity function (GLF) of the blazar population. Our previous GLF model is used as a basis, which features luminosity-dependent density evolution implied from X-ray data on active galactic nuclei, as well as the blazar sequence paradigm for their spectral energy distribution, and which is consistent with EGRET and current {\it Fermi} observations of blazars. Here we augment the high-redshift evolution of this model by utilizing the luminosity function of quasars from the Sloan Digital Sky Survey (SDSS), which is well-constrained up to $z\sim5$. We find that {\it Fermi} may discover a few blazars up to $z\sim6$ in the entire sky during its 5-year survey. We further discuss how such high-redshift blazar candidates may be efficiently selected in future {\it Fermi} data.
\end{abstract}

\begin{keywords}
galaxies : active -- gamma rays : theory.
\end{keywords}

\section{Introduction}
High-energy gamma-ray astronomy has progressed drastically after the launch of the Fermi Gamma-ray Space Telescope ({\it Fermi}). Its Large Area Telescope (LAT) is currently observing the entire gamma-ray sky in the 0.02 -- 300 GeV energy range  \citep{atw09}. The majority of extragalactic sources detected by {\it Fermi} are blazars, a subclass of active galactic nuclei (AGNs) dominated by broadband, nonthermal emission arising from relativistic jets oriented close to our line of sight \citep{abd09a,abd09b,abd10_catalog,abd10}. The {\it Fermi} 11 month catalog reports the detection of $\sim600$ blazars up to $z\sim3$ at Galactic latitudes $|b|>10^\circ$\citep{abd10}. It is expected that {\it Fermi} will detect more than 1000 blazars in the near future \citep[e.g.][]{nar06,der07,it09}. 

Before the {\it Fermi} era, the Energetic Gamma-Ray Experiment Telescope (EGRET) on board the Compton Gamma Ray Observatory detected $\sim$50 blazars in total up to $z\sim3$ \citep{har99} \footnote{\citet{rom04} and \citet{rom06} have reported one EGRET blazar candidate at $z\sim5.48$, but it has not been confirmed so far by {\it Fermi}. }. On the other hand, by analyzing the {\it Swift}/BAT blazar sample in the redshift range $z=$ 0.03 -- 4.0, \citet{aje09} have recently suggested that the density evolution of luminous blazars peak at $z= 4.3$. Since {\it Fermi} has an order of magnitude better sensitivity and positional accuracy at high Galactic latitudes compared to EGRET \citep{atw09}, it is naturally expected that {\it Fermi} will see much deeper into the universe in gamma-rays.

The purpose of this paper is to discuss expectations for the highest redshift blazars that {\it Fermi} may discover. This requires reasonable knowledge of their gamma-ray luminosity function (GLF). The blazar GLF has been discussed from different perspectives in many papers so far \citep{pad93,ste93,sal94,chi95,ste96,chi98,muc00,nar06,der07,it09}. \citet[hereafter IT09]{it09} and \citet[][hereafter ITM10]{itm10} have recently developed a new model of the blazar GLF that accounts for the blazar spectral sequence, as well as luminosity-dependent density evolution implied from the AGN X-ray luminosity function (XLF), and which is consistent with the EGRET and current {\it Fermi} data (see \S \ref{sec:glf}). However, the ITM10 GLF is uncertain for $z>3$, because the current observed number of X-ray AGNs and gamma-ray blazars above $z\sim3$ is insufficient to strongly constrain the model. On the other hand, the optical luminosity function (OLF) of AGNs has been well constrained up to $z\sim5$ \citep{ric06}. Therefore, here we consider a modified blazar GLF by combining the AGN XLF with the evolutionary constraints from the AGN OLF data.

This paper is organized as follows. We introduce our updated GLF model in \S \ref{sec:glf}. In \S \ref{sec:high_z}, we show predictions and candidate selection methods for high-redshift blazar observations with {\it Fermi}. A discussion and summary is given in \S \ref{sec:sum}. Throughout this paper, we adopt the standard cosmological parameter set ($h,\Omega_M,\Omega_\Lambda$)=(0.7,0.3,0.7).

\section{High Redshift Evolution of the Blazar Gamma-ray Luminosity Function}
\label{sec:glf}
\subsection{Blazar Gamma-ray Luminosity Function}
\label{subsec:glf}

IT09 recently developed a model for the blazar GLF featuring the so-called luminosity-dependent density evolution (LDDE),
based on the latest determination of the AGN XLF that clearly show the tendency of the more luminous objects to undergo their peak activity periods at higher redshifts \citep{ued03,has05}. Another novel aspect of IT09 was the account of the blazar sequence, which refers to the observed trend whereby the two characteristic frequencies at where the blazar spectral energy distribution (SED) peaks systematically decrease as the bolometric luminosity increases (\cite{fos97,fos98,kub98,don01,ghi09}, but see also \cite{pad07}). The key parameters in the GLF model have been carefully determined to match the observed flux and redshift distribution of EGRET blazars by a likelihood analysis. Although the blazar sequence SED in IT09 was observationally constrained only up to the EGRET band of 30 GeV, this was extended to include the TeV band in ITM10, using published TeV blazar data. In this paper, we use the ITM10 GLF as our baseline model. Since the main modification of the blazar sequence SED in ITM10 was for $\gtrsim100$ GeV, predictions for {\it Fermi} by ITM10 are similar to those by IT09 (see ITM10 for details). 

Including some contributions from radio-quiet AGNs \citep{itu08}, the extragalactic gamma-ray background (EGRB) spectrum predicted by IT09 was found to be consistent with that actually observed by {\it Fermi} \citep{ack10}. Furthermore, the expected number count for {\it Fermi} blazars was $\sim750$ all-sky for the typical 1-year survey sensitivity limit of $F(> \rm100\  MeV)=3\times10^{-9}\ \rm photons \ cm^{-2}\ s^{-1}$ \citep{atw09}. This also seems to be consistent with the numbers reported in the {\it Fermi} 11-month AGN catalog, which are 596 blazars and 72 unidentified gamma-ray sources at $|b|>10^\circ$. However, a very recent analysis of the {\it Fermi} extragalactic source distribution revealed that the surface density of blazars is $\sim$0.12 deg$^{-2}$ down to $F(> {\rm100\  MeV}) =1\times10^{-9}\ \rm photons \ cm^{-2}\ s^{-1}$, which would imply $\sim5,000$ blazars in the entire sky \citep{abd10_extra}. Although this is $\sim8$ times larger than the actual number of {\it Fermi} blazars, the difference may arise from corrections for the detection efficiency of the LAT. The discrepancy from our prediction is a factor of $\sim$4 down to this sensitivity. One possible source of the discrepancy is differences in the adopted range of gamma-ray luminosities, since our GLF is based on the EGRET sample that contained only $\sim10$ low-luminosity BL Lac objects \citep[see][]{abd10_extra}. Another reason for the discrepancy may lie in the correction factors for the detection efficiency, which could be as large as 100-1000 near the sensitivity limit, according to \citet{abd10_extra}. More detailed analyses of the {\it Fermi} data including information on the distributions of luminosity and redshift is necessary for clarification. If the true number of sources is greater than that in our model, it would only increase the prospects for finding high-redshift blazars with {\it Fermi}.

\subsection{Constraints on the High Redshift Evolution of Blazars}
\label{subsec:high_z}

\begin{figure*}
\centering
\includegraphics[width=128mm]{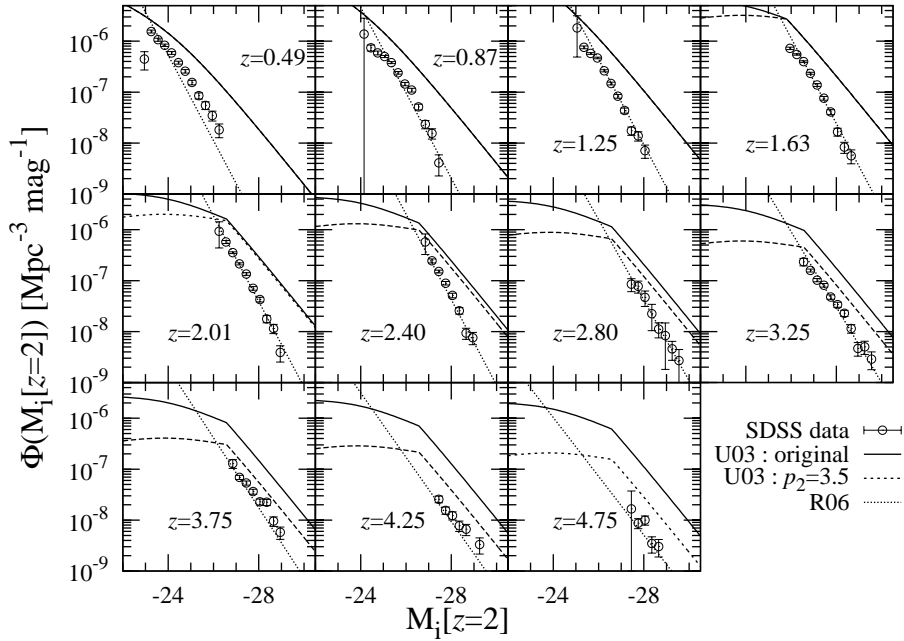}
\caption{$M_i [z=2]$ luminosity function at each redshift as indicated in each panel. Circles indicate SDSS quasar OLF data \citep{ric06}. Dashed, solid, and dotted curves show the U03 original XLF ($p_2=1.5$), modified U03 XLF ($p_2=3.5$), and R06 OLF, respectively.}
\label{fig:olf}
\end{figure*}

The ITM10 GLF model is based on data from EGRET blazars \citep{har99} and X-ray AGNs \citep{ued03,has05}, the highest redshifts for both samples being $z\sim3$. To address the evolution at $z \ge 3$, additional observational constraints are necessary. Optical surveys such as the Sloan Digital Sky Survey (SDSS) have successfully identified quasars up to $z=6.43$ \citep{fan03}, and the AGN OLF is well determined up to $z\sim5$ \citep[][hereafter R06]{ric06}. Utilizing the R06 OLF, below we modify the high-redshift evolution of our previous best-fit blazar GLF based on the AGN XLF of \citet[][hereafter U03]{ued03}.

The AGN XLF and OLF are merged following the procedures described in \S 5.4 of \citet{ric05} and \S 3, 4 in U03, which we briefly summarize. First, since the R06 OLF concerns optically selected type-I AGNs only while the U03 XLF includes all types of AGNs, the latter is converted as follows. From Fig. 9 in U03, the fraction of optically selected type--I AGNs as a function of absorption column density $N_{\rm H}$ can be characterized as
\begin{eqnarray}  
\lefteqn{ \zeta(N_{\rm H})}\nonumber \\
 &=\left\{\begin{array}{ll}
      0.96 & (20.0 \le \log N_{\rm H} < 20.5), \\
	8.504 - 0.368\log N_{\rm H} & (20.5 \le \log N_{\rm H} < 23.0), \\
	0.04 & (23.0 \le \log N_{\rm H} < 24.0). 
    \end{array}\right.
\end{eqnarray} 
From Eqs. 8--10 in U03, the probability distribution of $N_{\rm H}$ for AGNs having luminosity $L_{\rm X}$ at redshift $z$ is
\begin{eqnarray}  
\lefteqn{\eta(N_{\rm H};L_{\rm X},z)}\nonumber \\
 &=\left\{\begin{array}{ll}
     2-\frac{5+2\epsilon}{1+\epsilon}\psi(L_{\rm X},z)&  (20.0 \leq \log N_{\rm H} < 20.5), \\
	\frac{1}{1+\epsilon}\psi(L_{\rm X},z) & (20.5 \leq \log N_{\rm H} < 23.0), \\
	\frac{\epsilon}{1+\epsilon}\psi(L_{\rm X},z) & (23.0 \leq \log N_{\rm H} < 24.0)., 
    \end{array}\right.
\end{eqnarray} 
where $L_{\rm X}$ is the 2--10 keV luminosity in units of erg/s, $\epsilon=$1.7,
and \begin{equation}  
\psi(L_{\rm X},z)={\rm min}\{\psi_{\rm max}, {\rm max}[0.47-0.1(\log L_{\rm X} - 44.0), 0])\},
\end{equation}
for which
\begin{equation}
\psi_{\rm max}=\frac{1+\epsilon}{3+\epsilon}.
\end{equation}
Then the fraction of optically selected type--I AGNs is
\begin{equation}
f_{\rm opt. type-I}(L_{\rm X},z)=\int_{20.5}^{24.0} dN_{\rm H} \ \zeta(N_{\rm H})\times  \eta(L_{\rm X},z;N_{\rm H})   .
\end{equation}

Next, we convert $L_{\rm X}$ to $M_i [z=2]$, the {\it K}-corrected {\it i}-band AB magnitude at $z=2$. First, we evaluate the 2 keV luminosity $L_{\rm 2 keV}$ in units of erg/s/Hz by assuming a photon index of $\Gamma=1.9$. This is then extrapolated to $L_{\rm 2500}$, the luminosity at 2500 \AA \, by solving the equations 
\begin{eqnarray}
\alpha_{\rm ox}&=&-0.137 \log L_{\rm 2500} + 2.638, \label{eq:alpha_ox}\\
\log L_{\rm 2keV}&=&\log L_{\rm 2500} + \alpha_{\rm ox} \log \left(\frac{\nu_{\rm 2 keV}}{\nu_{\rm 2500}}\right),\label{eq:lum_ox}
\end{eqnarray}
where $\alpha_{\rm ox}$ is the spectral energy index between 2500 \AA\ to 2 keV \citep{ste06},
and $\nu_{\rm 2 keV}$ and $\nu_{\rm 2500}$ are respectively 2 keV and 2500 \AA \ in units of Hz.
Finally, we convert from luminosity $L_{\rm 2500}$ to magnitude $M_i[z=2]$ as \citep{oke83},
\begin{equation}
M_i[z=2] = -2.5\log\frac{L_{2500}}{4\pi d^2} - 48.60 - 2.5 \log(1+2),
\end{equation}
where $d=10 \ {\rm pc}=3.08\times10^{19} \ {\rm cm}$. Thus the U03 XLF can be combined with the R06 OLF.

The power-law index $p_2$ in the U03 XLF characterizes the density evolution as a function of $z$ at high redshift. Although $p_2=1.5$ was used in ITM10, here we change this to $p_2=3.5$ in order to be consistent with the high-redshift R06 OLF data. Since $p_2$ does not affect the low-redshift evolution, the XLF is not significantly altered below $z\sim3$. Fig. \ref{fig:olf} shows the AGN OLF at each redshift in terms of $M_i [z=2]$ in comparison with the R06 OLF data. With this AGN XLF, we reconstruct the blazar GLF with the ITM10 blazar sequence SED model, following the procedures of IT09. 

Note that there are some discrepancies between the OLF data and our OLF model at the brightest luminosities for low redshifts. This might be due to insufficient accuracy in our method of converting luminosities and luminosity functions. The correlation of Eqs.\ref{eq:alpha_ox} and \ref{eq:lum_ox} is derived from a sample of 293 AGNs \citep{ste06}, and the conversion between the XLF to OLF is based on a sample of only 247 objects \citep{ued03}. In order to reliably apply our model to wider ranges of luminosities and redshifts, the statistical uncertainties must be decreased through more data from deeper, larger X-ray and optical AGN surveys. 

The key parameters of our new blazar GLF model are $(q,\ \gamma_1, \kappa)=(4.42, 1.07, 1.92\times10^{-6})$,
where $q$ is the ratio between the bolometric jet luminosity and nuclear X-ray luminosity, $\gamma_1$ is the faint-end slope index of the GLF, and $\kappa$ is a normalization factor for the GLF (see Section. 3 of IT09 for details). Based on this model, we now discuss the prospects for observing high-redshift blazars with {\it Fermi}.

\section{Predictions for high-redshift Fermi blazars}
\label{sec:high_z}

\subsection{Expected Redshift Distribution of Fermi Blazars}

\begin{figure}
  \begin{center}
  \centering
  \includegraphics[width=80mm]{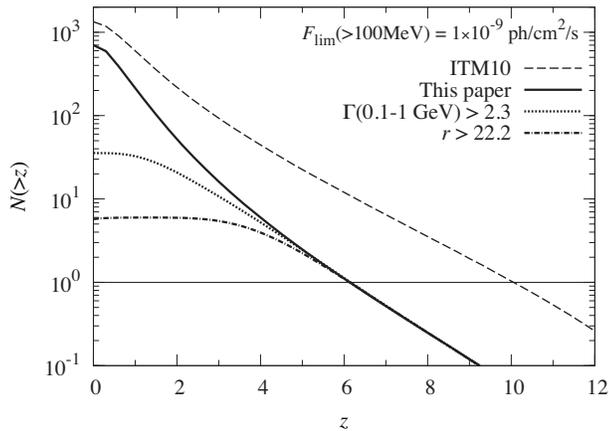}
\caption{Expected cumulative redshift distribution of blazars detectable by {\it Fermi} above 100 MeV in the entire sky for a flux sensitivity limit $F(>100{\rm MeV})=1\times 10^{-9} {\rm photons/cm^2/s}$. Solid and dashed curves correspond to whether or not AGN OLF constraints are taken into account. The dotted curve is for $\Gamma>2.3$, and the dot-dashed curve is 
for $\Gamma>2.3$ together with $r > 22.2$}\label{fig:z_dist}
  \end{center}
\end{figure}

Fig. \ref{fig:z_dist} shows the expected cumulative redshift distribution of {\it Fermi} blazars above 100 MeV in the entire sky for a $\sim$5-year survey flux sensitivity limit of $F(>100{\rm MeV})=1\times 10^{-9} \rm \ photons/cm^2/s$. The effects of intergalactic absorption due to $\gamma\gamma$ interactions with diffuse background radiation fields are not included here, as they are not expected to be important below 1 GeV \citep[e.g.][]{gil09,sin09}. Neglecting high-redshift optical constraints, {\it Fermi} may be able to detect blazars up to $z\sim10$. Taking into account their high-redshift evolution implied from the AGN OLF, we expect that {\it Fermi} will find some blazars up to $z\sim6$ with the $\sim$5-year survey sensitivity.

\subsection{Methods for selecting high-redshift blazar candidates}

It is expected that {\it Fermi} will find more than 1,000 blazars in total \citep{nar06,der07,it09}. Although $\sim 70$\% of the {\it Fermi} blazars have confirmed redshifts, the other $\sim 30$\% still do not \citep{abd10_catalog}. Furthermore, 10\% of the high-latitude {\it Fermi} sources remain unidentified \citep{abd10}.
This implies that the highest redshift blazars must be searched for among a large number of sources and distinguished from numerous unrelated ones. The following methods for their observational selection should be effective for this purpose.

First, we select the sources whose fluxes are close to the {\it Fermi} sensitivity limit, as the high-redshift blazars are naturally expected to be faint. Furthermore, it should be useful to look for those showing time variability, a characteristic trait of blazars \citep{urr95}, although this may not always be practical due to the faint fluxes.

Second, we choose the sources whose 0.1--1 GeV photon indices $\Gamma> 2.3$ to pick out blazars with gamma-ray luminosities $L_\gamma>10^{47}$ erg/s at 100 MeV, since high-redshift objects that are detectable should be luminous,
and their spectral indices are expected to be soft according to the blazar sequence \citep{fos98,kub98}. Although this criterion depends entirely on the assumed blazar sequence paradigm which is still a matter of debate, the above trend is already seen in the 3-month bright {\it Fermi} blazar catalog \citep{ghi09}.

Third, we search for the radio counterpart using deep radio survey catalogs, which should give us tighter constraints on the source location and identification. According to the blazar sequence, the radio fluxes of blazars at $z \sim 6$ would be $\sim20$ mJy, fainter than the limiting sensitivity of radio-loud galaxy catalogs used for the current {\it Fermi} catalog such as the Combined Radio All-sky Targeted Eight GHz Survey \citep[CRATES;][]{hea07}, the Candidate Gamma-Ray Blazar Survey \citep[CGRaBS;][]{hea08}, and the Roma-BZCAT \citep{mas09}. More useful for our purposes are the Faint Images of the Radio Sky at Twenty centimeters survey \citep[FIRST;][]{whi97} and the NRAO VLA Sky Survey \citep[NVSS;][]{con98}, which reach down to 1 mJy at 1.4 GHz in the northern sky. Within the $\it Fermi$ localization uncertainty of $\sim10'$ at $|b|>10^\circ$, the average radio source count is nominally expected to be $\sim0.20$ blazars, so finding the radio counterpart should be a crucial discriminant.

Fourth, we select objects that are not detected in optical surveys such as the SDSS, in view of the unavoidable attenuation due to intergalactic HI for high-redshift sources. Since FIRST covers the SDSS survey area, a quarter of the entire sky, sources detected by the SDSS can be searched for their radio counterparts. Here we set the optical selection criterion to be {\it r} $>22.2$ in AB magnitude, the {\it r}-band limiting magnitude of SDSS. We caution that the SDSS sensitivities in the {\it i} and {\it z} band would be insufficient for detecting blazars at $z>3$ even in the absence of intergalactic absorption. Thus, the SDSS data by itself cannot be used for identifying high-redshift blazars, but will nevertheless be valuable for rejecting low-redshift contaminants.

The dotted and dot-dashed curves in Fig. \ref{fig:z_dist} show cases where the above selection criteria have been imposed for the gamma-ray photon index and the optical magnitude. The expected total number of {\it Fermi} blazars would decrease to $\sim60$ sources, of which $\sim$38\% is expected to lie at $z>5$, so this procedure should be effective in narrowing down high-redshift blazar candidates, albeit with some remaining low-redshift contamination. Finally, deeper optical and infrared follow-up observations of the candidate objects are warranted to accurately identify their respective counterparts and spectroscopically measure their actual redshifts. Note that for blazars in the southern sky, surveys such as that being undertaken by the South Pole Telescope at 1.4 and 2.0 mm with milli-Jansky sensitivities \citep{vie10}, as well as the VST ATLAS \footnote{VST ATLAS: http://astro.dur.ac.uk/Cosmology/vstatlas/} and VISTA \footnote{VISTA: http://www.vista.ac.uk/} surveys in the optical and near-infrared, should prove to be likewise valuable for their selection. Even more powerful would be future all-sky near-infrared surveys by satellite missions such as the Joint Astrophysics Nascent Universe Satellite \citep[JANUS;][]{rom10}, Wide-field Imaging Surveyor at High Redshift (WISH)\footnote{WISH: http://www.wishmission.org/en/index.html}, or the Wide-Field Infrared Survey Telescope \citep[WFIRST;][]{ste10,geh10}.

\section{Discussion and Summary}
\label{sec:sum}
In this paper, we studied the prospects for detecting the highest-redshift blazars with {\it Fermi} through a model for the blazar GLF, and how we can select such blazars among numerous other expected {\it Fermi} sources. The high-redshift evolution of our GLF model was constrained from the observed AGN OLF, which is well determined up to $z\sim5$. Thus we found that {\it Fermi} may discover some blazars up to $z\sim6$ down to the $\sim$5-year survey flux sensitivity limit of $F(>100{\rm MeV})=1\times 10^{-9} \rm \ photons/cm^2/s$. If the true number of sources turns out to be larger than the predictions of our basic model (IT08,ITM09), the number of high-redshift blazars should be correspondingly larger.

High-redshift blazar candidates may be selected through the criteria that the source
(1) is faint and shows time variability,
(2) has a soft spectrum with $\Gamma>2.3$, corresponding to luminous objects in the blazar sequence,
(3) has a radio counterpart in deep survey catalogs such as FIRST, and
(4) are not detected in optical surveys such as SDSS, which should remove low-redshift contaminants.
Then these sources must be followed up by optical and infrared telescopes.

GeV photons from high-redshift sources are expected to suffer intergalactic absorption due to electron-positron pair production interactions with the cosmic UV background radiation. The resulting attenuation features in the spectra of such sources could therefore be a key probe of the poorly-understood UV background and its evolution, providing valuable insight into the epoch of cosmic reionization and early star formation \citep{oh01,gil09,sin09}. Discovering high-redshift gamma-ray blazars will constitute a crucial step towards this goal.

The typical isotropic bolometric luminosity $L_{\rm bol}$ expected from the blazar GLF is $10^{48}$ erg/s. If luminous blazars with $L_{\rm bol} > 10^{49}$ erg/s are found at $z \gtrsim 6$, the implied mass of the central supermassive black hole would be $M_{\rm BH} > 10^9 M_{\odot}$, assuming Eddington-limited jet power and typical jet Doppler factors $\sim10$. Given the difficulty of explaining the existence of such massive black holes at $z \gtrsim 6$ with Eddington-limited accretion initiated from stellar-mass seeds \citep{tana09,vol10}, future constraints on high-redshift luminous blazars may also provide important constraints on the growth history of supermassive black holes.

\section*{acknowledgments}
We thank J. D. Vieira for useful comments.
This work was supported in part by
the Global COE Program ``The Next Generation of Physics, Spun from
Universality and Emergence'' at Kyoto University from the Ministry of Education, Culture,
Sports, Science and Technology (MEXT) of Japan. YI acknowledges
support by the Research Fellowship of the Japan Society for the
Promotion of Science (JSPS),
and SI acknowledges support by the Grant-in-Aid No.22540278 from MEXT of Japan.
\bibliographystyle{mn2e} 
\bibliography{draft}  

\label{lastpage}
\end{document}